\documentclass[preprint2]{aastex6}
\usepackage{graphicx}
\usepackage{natbib}
\usepackage{amsmath}

\newcommand{\Msolar}{M$_{\odot}$}

\newcommand{\Rsolar}{R$_{\odot}$}
\newcommand{\kms}{km s$^{-1}$}

\shorttitle{Blue Straggler Spindown}
\shortauthors{Leiner et al.}

\begin{document}

\title{Observations of Spin-down in Post-Mass-Transfer Stars and the Possibility for Blue Straggler Gyrochronology}

\author{Emily Leiner\altaffilmark{1,2, 3}, Robert
  D. Mathieu\altaffilmark{2}, Natalie M. Gosnell \altaffilmark{4}, Alison Sills\altaffilmark{5}}
\email{emily.leiner@northwestern.edu}

\altaffiltext{1}{Center for Interdisciplinary Exploration and Research in Astrophysics, Northwestern University, 2145 Sheridan Rd, Evanston, IL 60208, USA}
\altaffiltext{2}{Department of Astronomy, University of Wisconsin-Madison, 475 North Charter St, Madison, WI 53706, USA}
\altaffiltext{3}{NSF Astronomy and Astrophysics Postdoctoral Fellow}
\altaffiltext{4}{Department of Physics, Colorado College, 14 E. Cache La Poudre St, Colorado Springs, CO  80903, USA}
\altaffiltext{5}{Department of Physics \& Astronomy, McMaster University, 1280 Main Street West, Hamilton, ON L8S 4M1, CANADA}

\begin{abstract}
Blue stragglers and other mass transfer/collision products are likely born with rapid rotation rates due to angular momentum transfer during mass-transfer, merger or collisional formation. However, less is known about the angular momentum evolution of these stars as they age. Here we compare rotation rates and post-formation ages of mass-transfer products to models of angular momentum evolution for normal main-sequence stars and collisionally formed blue stragglers. In our sample, we include both F- and G-type blue stragglers in the cluster NGC 188 and post-mass-transfer GK main-sequence stars in the field, all binaries with WD companions. We compare ages derived from WD cooling models to photometric rotation periods and/or spectral $v$sin$i$ measurements.  We demonstrate that these systems have rapid rotation rates soon after formation. They then spin down as they age much as standard solar-type main-sequence stars do. We discuss the physical implications of this result, which suggests that the spin-down of post-mass transfer stars can be described by standard magnetic-braking prescriptions. This opens up the possibility of using gyrochronology as a method to determine the time since formation of blue straggler stars and other post-mass-transfer binaries. 
\end{abstract}

\section{Introduction}\label{section:intro}
 In color-magnitude diagrams (CMDs) of star clusters, blue straggler stars (BSSs) are found brighter and bluer than the main sequence turnoff. BSSs are thought to form from mass transfer in binary systems \citep{McCrea1964, Gosnell2014}, stellar collisions during dynamical encounters \citep{Leonard1989, Sills2001}, or binary mergers, for example induced by Kozai cycles \citep{Perets2009}.

Main-sequence (MS) stars that have been through mass transfer or a merger also exist in the field. These BSS analogs can be identified by abundance anomalies -- e.g. barium stars, carbon enhanced metal poor stars (CEMPs), lithium enhanced giants  \citep{Jorissen1998,Hansen2016, Aoki2008}. In other cases, these stars are identified as blue stars with low metallicities indicative of an older population \citep{Preston2000}. Post-mass-transfer systems can also be identified by the direct detection of hot white dwarf (WD) companions to MS stars, often in UV surveys (e.g. \citealt{Holberg2013, Rebassa-Mansergas2017}). All of these objects are related classes of post-mass-transfer or post-merger binaries.

Mass transfer also transports angular momentum, resulting in spin-up of the mass-accreting star \citep{Packet1981, DeMink2013, Matrozis2017}. Similarly, stellar collisions and mergers are expected to yield rapidly rotating stars \citep{Sills2001, Sills2005}. These interactions can be seen as resetting the gyro-age clock, giving old stars the rapid rotation rates indicative of youth.  
 
 While little work has been done to compare observed rotation rates in post-mass-transfer systems to these theoretical predictions, observations do confirm qualitatively that many mass-transfer and collision products like the BSSs are rotating rapidly (e.g. \citealt{Carney2005, Jeffries1996, Mucciarelli2014, Lovisi2010}), sometimes with $v$sin$i$ measurements as large as 200 \kms. 

Less studied is how these stars spin down once mass transfer has ended.  Normal solar-type stars have long been known to spin down as they age \citep{Skumanich1972, Kraft1967}. Recently, the \textit{Kepler} mission has delivered rotation periods for thousands of MS stars and ushered in a new era of precision rotation studies (e.g. \citealt{Meibom2015, Meibom2011, Angus2015, McQuillan2014}). These studies show that solar-type stars begin their lives with a wide range of rotation periods. As they age, they spin down due to magnetic braking, with faster rotators spinning down more quickly due to their stronger magnetic fields. After several hundred Myr, solar-type stars of the same age will converge to the same rotation rate regardless of their initial angular momentum. This fact enables rotation rate to be used as an indicator of stellar age, a technique known as gyrochronology. 

Here we ask: do late-type MS stars in post-mass-transfer binaries spin down in the same way, and are gyrochronology ages useful proxies for the time since mass transfer ended in these systems? We seek to answer this question by assembling a sample of post-mass-transfer binaries with measured rotation rates and ages from white dwarf cooling models. We compare these results to spin-down models for single solar-type stars and for collisionally produced BSSs. Finally, we discuss how these results illuminate the physics of the mass-transfer process and the applicability of gyro-ages to post-mass-transfer binaries.

 \section{Ages and Rotation Periods for a Sample of Post-Mass-Transfer Binaries}\label{section:rotationalevolution}
 
 \subsection{The Sample}\label{thesample}
 
 \begin{table*}
    \centering
    \begin{tabular}{ccccccc}
    \hline
     ID & Source &  MS/BSS & WD T$_\mathrm{eff}$ & WD $\log g$ & WD age & P$_\mathrm{rot}$ \\
        &     &  Spectral Type          &    (K)    &                  &           (Myr)   & days\\
    \hline 
     WOCS 5379$^\mathrm{{a}}$& Gosnell et al. 2018 & F7V & $15400^{+280}_{-250}$ & $7.5^{+0.06}_{-0.05}$ & $230^{+40}_{-30}$ & $>$ 2.5\\            
     WOCS 4540$^\mathrm{{a}}$& Gosnell et al. 2018 & F6V & $17100^{+150}_{-100}$ & $7.7^{+0.04}_{-0.02}$ & 95$^{+7}_{-5}$ & 1.8$^{2.3}_{0.92}$\\
     WOCS 4348& \citet{Gosnell2015} & F5V & $13000\pm500$ & \textit{7.8} & $245^{+30}_{-25}$ & 1.2$^{1.5}_{0.6}$\\
     WOCS 5350& \citet{Gosnell2015}& F5V & $13200\pm500$ &  \textit{7.8} & $235^{+30}_{-25}$ & 5.3$^{7.1}_{2.9}$\\
     WOCS 1888& \citet{Gosnell2015}& F6V& $11200\pm500$ &  \textit{7.8} & $370^{+50}_{-40}$ & 3.6$^{4.7}_{1.9}$\\
     WOCS 2679& \citet{Gosnell2015}& F6V & $11300\pm500$ &  \textit{7.8} & $360^{+50}_{-40}$ & 1.4$^{1.9}_{0.8}$\\
     WOCS 4230& \citet{Gosnell2015}& F8V& $11800\pm500$ & \textit{7.8} & $320^{+40}_{-35}$  & 1.0$^{1.3}_{0.6}$\\
     RE 0044+09& \citet{Kellett1995} & K2V & $28700\pm1500$ & 8.41 & $51^{+13}_{-12}$ & 0.4 \\
     KOI-3278& \citet{Kruse2014} & GV & $10000\pm750$ & 8.14 & $840^{+220}_{-160}$  & 12.5  \\
     KIC 6233093 & \citet{Kawahara2018} & GV & $<10000$ & 8.0 & $>1000$  &  17.1 \\
     2RE J0357+283& \citet{JBR1996} & K2V & $35000\pm5000$ & 8.0 & $6.3^{+2.9}_{-2.3}$  & 0.4 \\
     HD 217411& \citet{Holberg2014} & K0V & $37200\pm300$ & 7.8 & $4.8^{+0.12}_{-0.12}$  & 0.6 \\
    \hline
    \multicolumn{7}{l}{$^\mathrm{a}$ WD atmosphere fits found assuming a cluster distance to NGC 188 of 1950 pc with a Plummer radius of 11 pc.}
    \end{tabular}
    \caption{Post-mass-transfer BSS (or MS)-WD binaries.}
    \label{tab:sample}
\end{table*}

 To provide a more robust comparison between the rotational evolution of BSSs and other post-mass-transfer systems to models of stellar angular momentum evolution, we assemble a sample of wide ($P_{orb} > 80$ days) post-mass-transfer binaries from the literature consisting of FGK-type primaries with detected WD companions. These WDs all have temperature measurements enabling age estimates for the post-mass-transfer systems from WD cooling models. The primaries in these systems also have rotational measurements from spot modulation or from spectroscopic $v$sin$i$ measurements.

 Our sample is composed of 12 binaries from the literature containing a WD and a BSS or MS star, all in close enough orbits to infer mass transfer would have taken place in their past, but not so close that current tidal effects would affect their rotation rates. In most cases, orbital periods have been measured for these systems from radial velocities or eclipses \citep{Gosnell2015, Geller2009, Kawahara2018, Kruse2014}. In a few cases, precise orbital periods are not known but constraints from radial velocities and/or astrometry indicate likely periods on the order of months or years \citep{Holberg2014, Kellett1995, Jeffries1996}.

 This sample includes photometric WD detections to BSSs in the old (7 Gyr) open cluster NGC 188 \citep{Gosnell2015}, extreme-UV detections of WD companions to field K-dwarfs \citep{Kellett1995, JBR1996, Holberg2014}, and Kepler detections of self-lensing binary systems containing WDs \citep{Kawahara2018, Kruse2014}. These varying detection methods allow us to span an age range from hot and young (detectable with EUV surveys), to intermediate age (requiring high-precision HST UV photometry), to quite old and cool (undetectable photometrically in binaries with solar-type primaries, but discovered in microlensing surveys).

 \subsection{White Dwarf Cooling Ages}
For uniformity, we adopt the WD temperature estimates for our sample from the literature, but determine our own WD cooling ages using the models of \citet{Holberg2006} and \citet{Tremblay2011}\footnote{\url{http://www.astro.umontreal.ca/~bergeron/CoolingModels}}, except for WOCS 5379 where we use the cooling models of \citet{Althaus2013} because the WD has a He-core instead of a CO-core (Gosnell et al. 2018, in prep). We adopt the $\log g$ values for the WDs from the literature when available. Five sources only have photometrically-detected WDs, and for those we assume a surface gravity of $\log g=7.8$, corresponding to an approximate WD mass of 0.5 $M_{\odot}$. These $\log g$ values are shown in italics in Table~\ref{tab:sample}. The ages are determined using a bilinear interpolation in $T_{\textrm{eff}}$ and $\log g$. 

In Table~\ref{tab:sample}, we list each source in our sample, along with the literature reference for the system, the primary's spectral type, the literature values for WD temperature and $\log g$ value, our age estimate from WD cooling, and the primary star's rotation rate (which we discuss in Section~\ref{rotationperiods}). This range in spectral types spans masses from $\sim$0.8 \Msolar~to 1.5 \Msolar.

\subsection{Rotation Periods}\label{rotationperiods}
The field stars in our sample have rotation-period measurements from photometric modulation. We adopt these from the literature source in Table 1, except in the case of the detection in \citet{Kawahara2018}, where we use the \citet{McQuillan2014} rotation period of 17.1 days measured from the \textit{Kepler} lightcurve. \citet{McQuillan2014} classify this as a marginal detection, but we confirm this detection with visual examination of the light curve. Our Lomb-Scargle periodograms \citep{Hartman2008} also confirm the $\sim17$ day period.

For BSSs in the cluster NGC 188 we have only $v$sin$i$ measurements from the spectral archive of the WIYN Open Cluster Study \citep{Geller2008}. For these systems, we convert the $v$sin$i$ measurement to a rotation period. First, we calculate photometric radii for the blue-straggler primaries. To do this, we adopt the temperatures from \citet{Gosnell2015}, V-band magnitudes from \citet{Sarajedini1999}, and a distance modulus and reddening for the cluster of $(m-M)_V$= 11.44 and E(B-V)= 0.09 \citep{Sarajedini1999}. 

We use this radius to convert the observed rotational $v$sin$i$  to a distribution of possible periods assuming a random, uniform distribution of possible inclinations. We adopt the median value of this period distribution in Figure~\ref{spindown}, and also show error bars corresponding to the interquartile range.

We report these periods in Table~\ref{tab:sample}.

\section{Post-Mass-Transfer Spin-down}\label{agerotrelationship}

\subsection{Observations}
\begin{figure*}[!phbt]
\centering
\includegraphics[angle=0, width= .8\linewidth ]{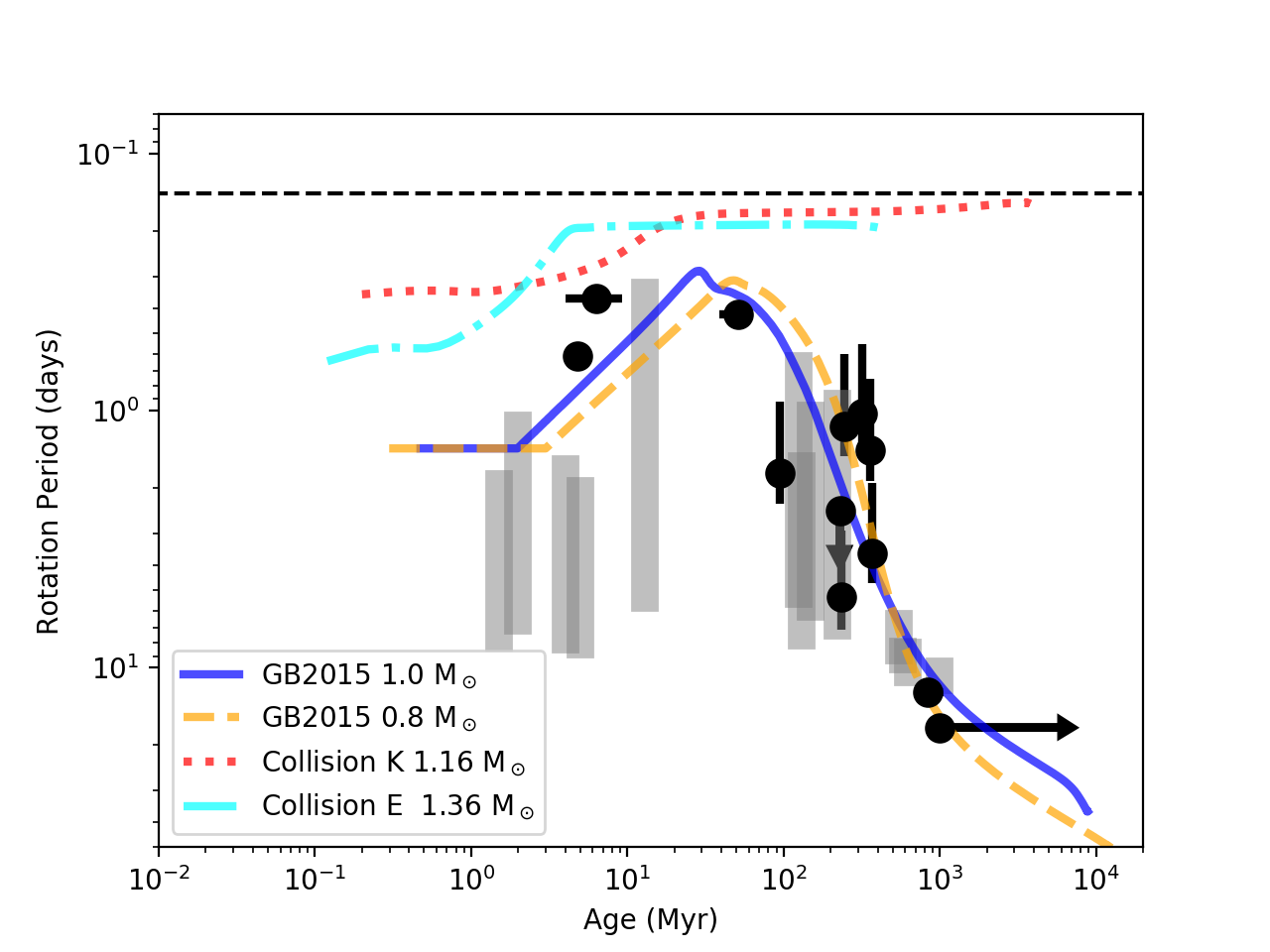}
\caption{We show the ages and rotation periods of our sample of post-mass-transfer binary systems. Arrows indicate where the WD age or rotation period is a limit. The gray bars indicate the range (25th percentile to 90th percentile) of observed rotation periods among $\sim 1$ \Msolar~MS stars in a sample of star clusters (taken from \citealt{Gallet2015} Table 1).  In cyan (dash dotted) and red (dotted) we show collision models from \citet{Sills2001}. In orange (dashed) and blue (solid), we show spin-down models for a from \citet{Gallet2015}. The dashed black line shows the critical rotation period for a 1.0\Msolar~, 1.0~\Rsolar~star \citep{Ekstrom2008}.}\label{spindown}
\end{figure*}

In Figure~\ref{spindown}, we show the relationship between the age (i.e. time since mass transfer ended) 
and the rotation period of the primary (i.e. the mass accretor) for the post-mass-transfer binary systems in our sample.The youngest post-mass-transfer stars in our sample ($t < 100$ Myr) are rotating with short periods of 0.4-0.6 days. The intermediate-aged systems of $100-400$ Myr have rotation periods ranging from about 1-10 days. The oldest stars in our sample with ages $> 660$ Myr have the slowest rotation periods of more than 10 days. 
 
The rotation periods match the spin-down models of \citet{Gallet2015} (hereafter GB2015) for single solar-type stars strikingly well. 

\subsection{Spin-down Models}
In Figure~\ref{spindown}, we compare our observations to two sets of models: 1) Spin-down models from GB2015 developed to match observed rotation rates of normal solar-type stars, and 2) spin-down models for BSSs formed from collisions between two MS stars. These two models offer two visions of angular momentum evolution in a post-mass-transfer stars: similar to typical spin-down on the MS, or dramatically different because of significant structural differences that might arise from stellar interaction. We briefly describe the evolution of these models below, and refer the reader to the original papers for further detail. 

\subsubsection{Gallet \& Bouvier 2015}
In Figure 1, we show spin-down models from GB2015 for 1.0 \Msolar~and 0.8 \Msolar.  The gray bars in Figure 1 show the distribution of rotation rates among $\sim1.0$ \Msolar~stars found in a sample of clusters of varying ages (GB2015, Table 1). The models we show are for ``fast rotators." They model the rotational evolution observed among cluster stars with rotation periods in the fastest $10\%$, and therefore follow the top of the cluster rotational distributions.  Briefly, the GB2015 models incorporate three physical processes: star-disk interaction during the pre-main-sequence (e.g. \citealt{Edwards1993, Matt2010}), angular momentum loss due to wind-driven magnetic braking on the MS (e.g. \citealt{Skumanich1972, Kawaler1988, Matt2012}), and redistribution of angular momentum within the stellar interior (e.g. \citealt{Spada2010, Eggenberger2017}). 

Models start with different initial rotation rates, with the ``fast rotator" model 
starting with $P_\mathrm{rot}= 1.4$ days, a rate matching the fastest decile of rotators in the youngest clusters. For the first few Myr the star is on the pre-main sequence with an accretion disk. The star's rotation is assumed to be locked to the disk, and thus the rotation rate remains constant for the duration of the disk lifetime ($\tau_\mathrm{disk}$). The star then contracts and spins up as it evolves to the MS, reaching the peak of the spin-down curve as it arrives on the zero-age main sequence. At this point, angular momentum loss due to wind-driven magnetic braking begins, and the star begins to spin down. The duration of this spin down is determined by the magnetic braking law used (in this case, \citealt{Matt2012}) along with an adopted scaling factor (K$_1$). During this time, too, it is possible for internal angular momentum redistribution to transfer angular momentum between the core and envelope. This time scale ($\tau_\mathrm{c-e}$) determines the shape of this spin down curve (see, for example, Fig. 4 in \citealt{Gallet2013}).

\subsubsection{Sills et al. 2001}

We also show two models in Figure~\ref{spindown} that model the rotational evolution of stellar collision products. In Case K (olive), two 0.6 \Msolar~ main-sequence stars collide. After a small amount of mass loss from the system, the final collision product is a 1.16 \Msolar~ star. In Case E (red), a 0.8~\Msolar~ and a 0.6~\Msolar~ star collide. After a small amount of mass loss, the final collision product is a 1.36 \Msolar~star. 

The initial collision products are bloated, luminous objects far from thermal equilibrium. They are also  rapidly rotating. As the stars contract back to thermal equilibrium, their large total angular momentum means that the stars reach break-up velocity almost immediately after the collision. As a result, \citet{Sills2001} chose to artificially reduce the initial angular momentum by a factor of 5, postulating that the angular momentum of these products must be quickly reduced after formation if the stars are to avoid completely disrupting. They do not specify the mechanism, but later work \citep{Sills2005} suggested that disk locking or some type of wind are both plausible mechanisms. 

During this early phase the luminosities of the collision products are powered by gravitational contraction. This phase lasts for $\sim10$ Myr, during which time the contracting stars spin up. The collision products then resume their lives on the MS, burning hydrogen for several hundred Myr or more with no significant angular momentum loss. 

\subsection{Physical Interpretation}\label{interpretation}

The spin-down of post-mass-transfer stars is remarkably similar to the spin-down models of normal main-sequence stars. 

It is not obvious why the angular momentum evolution of mass-transfer products should match the spin-down behavior of standard MS stars. For example, mass transfer might alter the mass accretor's structure and interior angular momentum distribution, or change the magnetic field strength or configuration. Here we discuss the physical implications of this result. 

First, let us consider the youngest stars in our sample with ages $< 100$ Myr. These stars have rotation rates of 0.4-0.6 days, corresponding to$~\sim30\%$ of break-up velocity. On one hand, rapid rotation is unsurprising as mass-transfer models predict substantial spin-up as a result of mass accretion \citep{Packet1981, DeMink2013, Matrozis2017}. On the other hand, it is notable that these stars are rotating at similar rates. Given that the youngest of these systems is $< 5$ Myr old, these stars have not had a chance to substantially spin down via the standard magnetized wind. Instead, whatever set the angular momenta of these objects must have occurred during or shortly after mass transfer. \citet{Packet1981} argue that accretion from a disk should be limited due to spin up, as the surface of an accreting star should reach Keplerian rotation after accreting just a few percent of its mass, preventing any further accretion onto the star. More recently, \citet{Matrozis2017} argue that given the masses observed among CEMPs and barium stars, these stars must be able to accrete several tenths of a solar mass of material from their AGB companions. To accrete so much mass, they argue, would require an efficient angular momentum loss mechanism to act during the mass transfer process -- perhaps disk locking or ejection of significant material through a strong wind -- effectively capping the star's angular momentum so that more material can be accreted. It seems plausible, then, that the maximum rotation rate for a post-mass-transfer star might be limited by such angular momentum regulation, and thus all three of our young systems are rotating near this maximum. Adding more young systems to this sample is necessary to further explore this idea and determine if  young FGK-type post-mass-transfer stars are indeed all rotating with similar periods, or if there is actually a larger spread in rotation rate than is evident in our small sample. 

The collisional models, in comparison, have slightly faster initial rotation rates, but we caution against reading too much into this comparison. Given the wide range of initial conditions and the rescaling of the initial angular momentum that has been applied to theses models, one could tune the collision models to have a wide range of rotation rates.

Notably, these 0.4-0.6 day rotation rates are comparable to the fastest rotation rates observed among solar-type stars in young clusters (gray bars in Figure~\ref{spindown}) and also to the the peaks in the GB2015 models. Perhaps angular momentum growth during pre-main-sequence accretion is similarly capped.

The older ($> 100$ Myr) systems in our sample allow us to explore spin-down behavior well after mass-transfer has ended. Here the spin-down rate is primarily determined by magnetic braking via a stellar wind (parameterized in the GB2015 models by a scaling factor $K_1$ to the \citet{Matt2012} law). The post-mass-transfer stars are therefore spinning down at the rate expected for their mass. A straightforward interpretation is that magnetic braking must be operating as usual in these stars, suggesting that these stars have normal convective envelopes and magnetic fields. 
In contrast, the \citet{Sills2001} collision models do not spin down via magnetic braking as they age. While typical stars in this mass range do spin down due to magnetic braking, the collision products are slightly hotter and brighter, and so never develop a convective envelope. Thus magnetic braking is never expected to take affect, and the stars maintain rapid rotation rates throughout their main-sequence evolution. 


The similarity of the GB2015 models and the spin-down behavior observed by our sample of FGK-type post-mass-transfer binaries suggests that gyrochronology relationships developed for standard solar-like stars are also applicable to post-mass-transfer systems. Rotation rates among post-mass-transfer systems like the BSSs, then, may be a useful proxy for time since formation, with recently formed systems rotating at a large fraction of brake up velocity, and older systems converging to rotation rates reflecting their age. Even in the absence of direct detections of white dwarf companions, then, comparing the rotation rate of a post-mass-transfer star to models like GB2015 could allow us to infer the time since the mass-transfer event. This possibility is particularly useful given the difficulty of detecting older, fainter white dwarf companions around FGK main sequence stars, and opens up a new method to determine formation rates and lifetimes for these systems, timescales that remain uncertain. 

Gyrochronology ages are generally not precise for young ($<1$ Gyr) main-sequence stars due to the large spread in their rotation rates at young ages. However, our post-mass-transfer sample has little scatter at young ages, suggesting gyrochronology may by a much more useful age-dating technique for young post-mass-transfer systems. 

While direct detection of a white dwarf companion is helpful in establishing that a star has a mass-transfer origin, there are many systems in which mass-transfer can be reasonably assumed without detecting the WD companion directly. Stars with abundance anomalies--e.g., barium stars or CEMP stars--are good examples. In addition, post-mass-transfer systems have distinctive orbital properties ($\sim$1000 day periods, near circular, mass functions indicating white dwarf companions; e.g. \citealt{Jorissen1998, Carney2005}). Many blue stragglers are found to have these orbital properties (e.g. \citealt{Gosnell2015}), and gyrochronology might reasonably be applied to any of these systems.

\section{Summary and Discussion}
In this paper, we compile a sample of long-period FGK-type MS stars with WD companions in the old open cluster NGC 188 and in the field. The orbital periods and presence of WD companions indicates that these systems are all post-mass-transfer binaries. We compare the ages of these systems from WD cooling models to the rotation periods of the FGK primaries. Our results show these post-mass-transfer systems are rotating at $\sim30\%$ of break-up at ages of 5-10 Myr, These periods are comparable to the fastest rotation rates observed among solar-type stars in young ($\sim10$ Myr) clusters. 

These post-mass-transfer stars then spin down as they age much as typical stars do. The spin-down behavior among stars in our sample agrees well with the spin-down models of GB2015, indicating that spin down could be occurring via a magnetized wind on comparable time scales to those found in normal stars. This result suggests that spin down behavior is not affected by whether a solar-type star was spun up on the pre-main-sequence or through a mass-transfer event. In both cases, the star seems to have the convective envelope and magnetic field required for wind angular momentum loss. 

Further, these results suggest that gyrochronology is a viable method to determine the time since formation of BSSs and other post-mass-transfer systems like barium stars, CEMPs, etc. 

This result can be refined by developing a larger sample of wide MS-WD binaries with known orbital solutions, rotation rates, and cooling ages. More young post-mass-transfer systems with hot WD companions are known from UV surveys (e.g. \citealt{Holberg2013}), but need to have rotation and orbital periods measured. Detecting older, cooler WDs with FGK companions is difficult, as these quickly become photometrically undetectable as they cool. 

In this regard we note that in addition to the BSSs included in the sample here, \citet{Gosnell2015} study 8 BSSs in NGC 188 with no detectable WDs. They argue that many of these systems likely also formed from mass transfer, but formed $> 400$ Myr ago so their WDs are too faint to detect. These BSSs are rotating slowly, with $v$sin$i$ $<$ 10 \kms \space (the resolution limit of the WOCS spectra), corresponding to rotation periods of longer than a few days. These limits on age and rotation rate are consistent with the hypothesis that these BSS are spinning down following the GB2015 models. 

With recent and future time-series photometric surveys like Kepler, TESS, and PLATO, our understanding of spin-down behavior several hundred Myr after mass transfer ends also will be advanced by more serendipitous detections of self-lensing WD-MS systems. 

These results are based on long-period (100s or 1000s of days) post-mass-transfer binaries with FGK main-sequence primaries. This behavior may not hold for higher-mass or lower-mass accretors or shorter-period binary systems that must have evolved through a common-envelope phase. In addition, BSSs may form in other ways, e.g. in dynamical collisions in clusters \citep{Leonard1989, Sills2001} or in binary mergers \citep{Perets2009, Andronov2006}. While these systems may also be rapidly rotating, their angular momentum evolution could be quite different \citep{Sills2001, Sills2005}. Indeed, the discovery of anomalously slow rotating A-type BSSs likely formed in collisions or mergers \citep{TakadaHidai2017, Fossati2010} offers evidence that the spin-down process for these stars may be much different. 

Our study suggests that all young post-mass-transfer stars are fast rotators, but our sample is small. If a larger spread in initial rotation velocities exists for post-mass-transfer stars, it remains to be seen if the slower rotators also spin down following typical models. 

With these limitations in mind, this work provides new insights into post-mass-transfer spin-down and provides the first evidence that rotation can be a useful clock for understanding formation timescales of post-mass-transfer objects. 

\acknowledgements{EL is supported by an NSF Astronomy and Astrophysics Postdoctoral Fellowship under award AST-1801937. RDM acknowledges funding support from NSF AST- 1714506 and NASA-NNX15AW69G. The authors thank Florian Gallet for providing models for this paper, and the anonymous referee for their helpful suggestions.}

 \end{document}